\documentclass[11pt]{article}

\usepackage{amsmath,dsfont}
\usepackage[english]{babel}
\usepackage{amsfonts}
\usepackage{amsmath,amssymb}
\usepackage{graphicx}

\makeatletter

\def\Z{\mathbb Z}

\def\eqnarray{\stepcounter{equation}\let\@currentlabel=\theequation
\global\@eqnswtrue
\global\@eqcnt\z@\tabskip\@centering\let\\=\@eqncr
$$\halign to \displaywidth\bgroup\@eqnsel\hskip\@centering
  $\displaystyle\tabskip\z@{##}$&\global\@eqcnt\@ne
  \hfil$\displaystyle{\hbox{}##\hbox{}}$\hfil
  &\global\@eqcnt\tw@ $\displaystyle\tabskip\z@
  {##}$\hfil\tabskip\@centering&\llap{##}\tabskip\z@\cr}
\@addtoreset{equation}{section}
  \def\theequation{\thesection.\arabic{equation}}
\makeatother

\usepackage{amsmath,amssymb}
\def\beq{\begin{equation}}
\def\eeq{\end{equation}}
\def\beqa{\begin{eqnarray}}
\def\eeqa{\end{eqnarray}}
\def\barray{\begin{array}}
\def\earray{\end{array}}

\begin{document}
%%%%%%%%%%%%%%%%%%%%%%%%%%%%%%%%%%%%%%%%%%%%%%%%%%%

\title{
{\bf
Self-isospectrality,
mirror symmetry, and exotic nonlinear supersymmetry
}}

\author{\bf Mikhail S. Plyushchay${}^{a,b}$ and
{Luis-Miguel
Nieto$^{b}$
}\\
[4pt]
{\small \textit{${}^{a}$ Departamento de F\'{\i}sica,
Universidad de Santiago de Chile, Casilla 307, Santiago 2,
Chile  }}\\
{\small
\textit{${}^{b}$ Departamento de F\'{\i}sica Te\'orica,
At\'omica y \'Optica, Universidad de Valladolid, 47071,
Valladolid, Spain}}\\
 \sl{\small{E-mails: mplyushc@lauca.usach.cl, luismi@metodos.fam.cie.uva.es
}}
}
\date{}

\maketitle

\begin{abstract}
We study supersymmetry
of a \emph{self-isospectral} one-gap
P\"oschl-Teller system
in the light of a mirror symmetry
that is based on
spatial and shift reflections.
The revealed exotic, partially broken nonlinear supersymmetry
admits seven alternatives for a grading operator.
One of its local, first order supercharges may be identified as a
Hamiltonian of an associated one-gap, non-periodic
Bogoliubov-de Gennes system.
The latter possesses a
nonlinear supersymmetric structure, in which
any of the three non-local generators of a Clifford
algebra may be chosen as the grading operator.
We find that the supersymmetry generators
for the both systems are the
Darboux-dressed integrals of a free spin-1/2 particle
in the Schr\"odinger picture,
or of a free massive Dirac particle.
Nonlocal Foldy-Wouthuysen transformations
are shown to be involved
in the supersymmetric structure.

\end{abstract}

\vskip.5cm\noindent

%%%%%%%%%%%%%%%%%%%%%%%%%%%%%%%%%%%%%%%%%%%%%%%%%%%%%%%%%%%%
%%%%%%%%%%%%%%%%%%%%%%%%%%%%%%%%%%%%%%%%%%%%%%%%%%%%%%%%%%%%
\section{Introduction}

A $\Z_2$ grading structure lies in the basis of supersymmetry.
In the early years of supersymmetric quantum mechanics~\cite{Witten,SUSYQM},
Gendenshtein and Krive observed~\cite{GK}
that in some systems the $\Z_2$ grading  may be provided
by a reflection operator. The origin of such a
hidden supersymmetric structure~\cite{SUSYbos,hidnon,CP1}
was explained recently in~\cite{JNP} by means of a Foldy-Wouthuysen
transformation for the case of a \emph{linear}
supersymmetry that is based on the
 \emph{first order} Darboux transformations~\cite{MatSal}
and is described by the Lie superalgebraic relations.

Braden and Macfarlane~\cite{BradMac}, and in a more broad context
Dunne and Feinberg~\cite{DF} revealed that a \emph{linear }N = 2
supersymmetric extension of the
\emph{periodic} finite-gap quantum systems
may produce completely isospectral systems characterized
by the same, but a shifted potential.
The name \emph{self-isospectrality}  was coined
by the latter authors for such a phenomenon, that was studied later
 by Fernandez et al~\cite{Fer+} as \emph{Darboux displacements},
 see also~\cite{FerGan}.

The both periodic and non-periodic \emph{finite-gap}
quantum systems, being
related to nonlinear integrable systems~\cite{FiniteGap},
find many important applications in diverse areas of physics,
ranging from condensed matter physics, QCD
and cosmology, to the string
theory~\cite{DHN,Bish,Fein,SchTh,
BasDun,CDP,FinCosm,Zwieb,DonWit,KriPho,BraMa,Str}.

A higher order generalization of the Darboux
transformations, known as the Darboux-Crum transformations~\cite{MatSal}, gives rise to
a higher derivative generalization of supersymmetric quantum mechanics~\cite{AIS},
characterized by \emph{nonlinear} superalgebraic
relations~\cite{hidnon,BagSam,FerSUSYn,KP1}.

Soon after the discovery of the self-isospectrality,
it was found that in some periodic
finite-gap systems this phenomenon
may be associated with not a linear, but nonlinear
supersymmetry~\cite{Fin3}. Later on,
hidden nonlinear supersymmetry~\cite{hidnon}
was revealed in unextended finite-gap
periodic finite-gap systems~\cite{CNP1}.
It was also established that self-isospectral
$n$-gap periodic systems with a half-period shift are described
by a special nonlinear supersymmetric
structure, that includes a hidden supersymmetry
of the order $2n+1$, whose local generator, being a Lax operator,
factorizes into the Darboux intertwining operators
of the explicit nonlinear, of order $2k$, $k\geq 1$,
 and linear or nonlinear,
of order $2(n-k)+1$,
supersymmetries~\cite{Tri}.

There is an essential difference between supersymmetries
of the periodic and non-periodic \emph{self-isospectral}
finite-gap systems.
In the former case, \emph{linear} $N=2$ supersymmetry generators,
as a part of a broader structure,
may annihilate two states of zero energy, while
they cannot have zero modes in the non-periodic case.
A little attention was
given, however, to  the study of the self-isospectrality
phenomenon in the non-periodic finite-gap systems.
\vskip0.08cm

In the present paper, we investigate the interplay of
the self-isospectrality,
reflections, Darboux transformations, and nonlinear
and hidden supersymmetries for
 \emph{non-periodic} finite-gap quantum systems.
This is done here for the simplest
case of a one-gap, self-isospectral reflectionless
P\"oschl-Teller (PT) system, and an associated one-gap
Bogoliubov-de Gennes (BdG) system
that is described by a first order Hamiltonian~\footnote{
The BdG system~\cite{BdG}  appears in many physical
problems, including, particularly, superconductivity theory,
fractional fermion number, the Peierls effect and
the crystalline condensates in the
chiral Gross-Neveu and Nambu-Jona Lasinio models,
see~\cite{DHN,Bish,Fein,SchTh,BasDun,CDP,GroNev,JacReb,Klim}.}.
We reveal a rich supersymmetric structure,
related to several admissible choices of the grading operator
(seven for PT and three for BdG) in these related systems.
Our analysis is based on a mirror symmetry
that includes a free particle as an essential element.
We find that all the nontrivial integrals are
a Darboux-dressed form
of the corresponding integrals of a free spin-1/2 particle system,
and show that nonlocal Foldy-Fouthuysen transformations are
involved in the exotic supersymmetric structure.
\vskip0.08cm

The paper is organized as follows.
In the next Section, a mirror symmetry of the self-isospectral,
 one-gap
reflectionless PT system is discussed, and its local and nonlocal
integrals of motion are identified
via Darboux dressing of a free particle.
In Section 3 we analyze
the eigenstates of the three basic local integrals.
Nonlinear superalgebraic structure and its peculiarities
are described  in Section 4.
In Section 5 we show that the unextended, single one-gap PT system may be
characterized by an exotic hidden nonlinear supersymmetry,
which is related
to supersymmetry of the extended, self-isospectral system by a nonlocal
Foldy-Wouthuysen transformation. In Section 6, identifying one of the
local supercharges of the self-isospectral PT system as
a $(1+1)D$ Dirac Hamiltonian,
we describe the nonlinear supersymmetry of the
associated one-gap BdG system.
Section 7 is devoted to the discussion of the results.

\section{Mirror symmetry and integrals of motion of self-isospectral one-gap
PT system}

Consider a one-gap, non-periodic  \emph{reflectionless}
P\"oschl-Teller (PT)
system~\cite{MatSal,GaiMat}~\footnote{The Hamiltonian of the
reflectionless one-gap system of the most general form is
$H_1=-d^2/dx^2-2\alpha^2\cosh^{-2}\alpha (x-x_0)+const$; we
put here $\alpha=1$, $const=1$, and fixed, for the moment,
$x_0=0$.},
\begin{equation}\label{PT}
    H_1=-\frac{d^2}{dx^2}-2\cosh^{-2}x+1\,,
\end{equation}
and factorize the Hamiltonian,
\begin{equation}\label{H1}
      H_1=A A^\dagger ,\qquad A=\frac{d}{dx}-\tanh x\,.
\end{equation}
It is connected with a (shifted for a constant) free
particle Hamiltonian,
\begin{equation}\label{H0}
    H_0=A^\dagger A=-\frac{d^2}{dx^2}+1\,,
\end{equation}
by the intertwining relations,
\begin{equation}\label{D1}
    AH_0=H_1A,\qquad
    H_0A^\dagger=A^\dagger H_1.
\end{equation}
The PT system (\ref{PT}) is almost isospectral to the
system (\ref{H0}).  The eigenstates of the same energy,
$H_1\psi_1^E=E\psi_1^E$, $H_0\psi_0^E=E\psi_0^E$, are
related~\footnote{Up to constant, energy-dependent
factors which are of no
importance for us here.} by a Darboux transformation
\begin{equation}\label{Darpsi}
    \psi_1^E(x)=A\psi_0^E(x),\qquad
    \psi_0^E(x)=A^\dagger\psi_1^E(x),
\end{equation}
 and  the spectra are
in-one-to-one correspondence except one bound, square
integrable state of zero energy, which is missing in the
free particle spectrum. Explicit form of the PT eingestates
is
\begin{eqnarray}\label{specsin}
    &E=0:\,\,\,\,\, \Psi^0(x)=\frac{1}{\cosh
    x}\,;\qquad\quad\quad
    E=1:\,\,\, \Psi^1(x)=
    -\tanh x\,;&\\
    &E=1+k^2>1:\,\,\,\,\, \psi^{\pm
    k}(x)=(\pm ik-\tanh x)e^{\pm ikx}\,,\,\,\,k>0.&\label{specdoub}
\end{eqnarray}
The \emph{doublet} states of the continuous part of the
spectrum ($E> 1$) are obtained from the plane wave states
$e^{\pm ikx}$, the \emph{singlet} state $\Psi^1$
corresponds to a singlet state $\psi_0^1=1$ ($k=0$) of the
free particle. A nonphysical state $\psi^0_0=\sinh x$,
which is a formal eigenstate of $H_0$, is mapped to the
unique  \emph{bound singlet} state $\Psi^0$ in the PT
system, $\Psi^0=A\psi^0_0$. The latter is a zero mode of
the first order operator $A^\dagger$, $A^\dagger\Psi^0=0$.
There is one energy gap in the spectrum of the
reflectionless PT system, that separates a zero energy
eigenvalue of the bound state from the continuous part of
the spectrum ($k\geq 0$). \vskip0.1cm

Let us shift the coordinate $x$ for $+\tau$ and for
$-\tau$ ($\tau>0$), and denote
$$
A_\tau=\frac{d}{dx} -\tanh (x+\tau)\,,\quad
A_{-\tau}=\frac{d}{dx}-\tanh(x-\tau)\,,\quad H_\tau=A_\tau
A^\dagger_\tau\,,\quad
H_{-\tau}=A_{-\tau}A^\dagger_{-\tau}\,.
$$
As the PT system $H_\tau$ is just the $H_{-\tau}$
translated for $2\tau$, these two Hamiltonians are
\emph{completely isospectral}.

 The
systems $H_\tau$ and $H_{-\tau}$ are related by a
\emph{mirror} (with respect to $x=0$) \emph{symmetry},
\begin{equation}\label{mirror1}
    \mathcal{R}H_\tau=H_{-\tau}\mathcal{R}\,,
\end{equation}
where $\mathcal{R}$ is a spatial reflection operator,
$\mathcal{R}x=-x\mathcal{R}$, $\mathcal{R}^2=1$.
 The reflection $\mathcal{R}$ intertwines therefore
the two isospectral PT systems, cf. (\ref{D1}). It also
intertwines the factorizing operators,
\begin{equation}\label{RA}
    \mathcal{R}A_\tau=-A_{-\tau}\mathcal{R}\,,\qquad
    \mathcal{R}A^\dagger_{\tau}=-A^\dagger_{-\tau}\mathcal{R}\,.
\end{equation}

In addition, we introduce a
reflection
operator~\footnote{From a viewpoint
of an associated free Dirac particle system, see below,
$\mathcal{T}$ may be treated as a
kind of a charge conjugation operator.}
for the shift parameter $\tau$,
$\mathcal{T}\tau=-\tau\mathcal{T}$, $\mathcal{T}^2=1$,
which also intertwines the Hamiltonians and the factorizing
operators,
\begin{equation}\label{IH}
    \mathcal{T}H_\tau=H_{-\tau}\mathcal{T}\,,
\end{equation}
\begin{equation}\label{IA}
    \mathcal{T}A_{\tau}=A_{-\tau}\mathcal{T}\,,\qquad
    \mathcal{T}A^\dagger_{\tau}=A^\dagger_{-\tau}\mathcal{T}\,.
\end{equation}

Each of the shifted Hamiltonians, $H_\tau$ and $H_{-\tau}$,
may also be treated as a mirror image of another, with a
free particle system playing the role of the mirror.
Indeed, a shift of $x$ does not change the free particle
Hamiltonian (\ref{H0}), $H_0=A^\dagger_\tau A_\tau=
A^\dagger_{-\tau} A_{-\tau}$, and we get the two different
sets of intertwining relations,
\begin{equation}\label{interbeta+}
    A_{\tau}H_0=H_\tau A_{\tau},\qquad
    H_0A^\dagger_{\tau}=A^\dagger_{\tau} H_\tau,
\end{equation}
\begin{equation}\label{interbeta-}
    A_{-\tau}H_0=H_{-\tau}A_{-\tau},\qquad
    H_0A^\dagger_{-\tau}=A^\dagger_{-\tau} H_{-\tau}.
\end{equation}
Combining them, we find the \emph{second order} operators
that generate a Darboux-Crum  transform between the two
mutually shifted PT systems,
\begin{equation}\label{HH}
    Y_{\tau}H_{\tau}=
    H_{-\tau}Y_{\tau}\,,\qquad
    Y_{-\tau}H_{-\tau}=
    H_{\tau}Y_{-\tau}\,,
\end{equation}
where
\begin{equation}\label{Y}
    Y_\tau=A_{-\tau}A^\dagger_{\tau}\,,\qquad
    Y^\dagger_\tau=Y_{-\tau}\,.
\end{equation}
The mirror $H_0$ is present virtually here by means of
relations (\ref{interbeta+}) and (\ref{interbeta-}),
\begin{equation}\label{mirrorH0}
Y_\tau H_\tau=A_{-\tau}(A^\dagger_\tau
H_\tau)=A_{-\tau}(H_0
A^\dagger_\tau)=(A_{-\tau}H_0)A^\dagger_\tau=(H_{-\tau}
A_{-\tau})A^\dagger_\tau=H_{-\tau}Y_{\tau}\,.
\end{equation}

The Darboux-Crum intertwining relations (\ref{HH}) are
translated into the language of supersymmetric quantum
mechanics. Consider the composed system described by the
diagonal two-by-two Hamiltonian
\begin{equation}\label{Hsusy}
    \mathcal{H}=\left(%
\begin{array}{cc}
  H_\tau & 0 \\
  0 & H_{-\tau} \\
\end{array}%
\right),
\end{equation}
and define the matrix operators
\begin{equation}\label{Q12}
    Q_1=\left(%
\begin{array}{cc}
  0 & Y_{\tau}^\dagger \\
  Y_{\tau} & 0 \\
\end{array}%
\right),\qquad Q_2=i\sigma_3 Q_1\,.
\end{equation}
Due to (\ref{HH}), the $Q_1$ and $Q_2$ are the integrals of
motion of the extended system (\ref{Hsusy}),
$[\mathcal{H},Q_a]=0$, $a=1,2$. The diagonal Pauli matrix
$\sigma_3$ can be taken as a grading operator,
$\Gamma=\sigma_3$, $\Gamma^2=1$. Then $\mathcal{H}$ and
$Q_a$ are identified, respectively, as bosonic and
fermionic operators, $[\Gamma,\mathcal{H}]=0$,
$\{\Gamma,Q_a\}=0$. With taking into account Eqs.
(\ref{interbeta+}), (\ref{interbeta-}), one finds that the
supercharges $Q_a$ generate a nonlinear, second order
superalgebra
\begin{equation}\label{QQ}
    \{Q_a,Q_b\}=2\delta_{ab}\mathcal{H}^2.
\end{equation}

 The system (\ref{Hsusy}), being a one-gap
(super-extended)  \emph{self-isospectral}
reflectionless system, possesses other nontrivial
integrals~\footnote{For earlier discussions of this system
see~\cite{Bish,SchTh,CDP,AS}.}.
To find them, we use the following observation
\cite{AdS}. Suppose that some
Hamiltonians~\footnote{Intertwined Hamiltonians $H$ and
$\tilde{H}$ can be Hermitian operators of any, including
matrix, nature.}, $H$ and $\tilde{H}$, are related by the
intertwining identities $DH=\tilde{H}D$,
$HD^\dagger=D^\dagger\tilde{H}$, where $D$ is a
differential operator of any order. If ${J}$ is an integral
of the system $H$, then $D{J}D^\dagger$ is the integral of
the system $\tilde{H}$,
\begin{equation}\label{JJ}
    [J,H]=0\,\,\,\Rightarrow\,\,\,
    [\tilde{J},\tilde{H}]=0,\quad
    \tilde{J}=DJD^\dagger\,.
\end{equation}
Associate with the system (\ref{Hsusy}) an extended system
\begin{equation}\label{Hsusy0}
    \mathcal{H}_0=\left(%
\begin{array}{cc}
  H_0 & 0 \\
  0 & H_0 \\
\end{array}%
\right),
\end{equation}
composed from the two copies of the free particle. The
systems (\ref{Hsusy}) and (\ref{Hsusy0}) are related, in
correspondence with (\ref{interbeta+}), (\ref{interbeta-}),
by the identities
$\mathcal{D}\mathcal{H}_0=\mathcal{H}\mathcal{D}$,
$\mathcal{H}_0\mathcal{D}^\dagger=
\mathcal{D}^\dagger\mathcal{H}$, where the matrix
intertwining operator is
\begin{equation}\label{Dmat}
    \mathcal{D}=\left(%
\begin{array}{cc}
  A_\tau & 0 \\
  0 & A_{-\tau} \\
\end{array}%
\right).
\end{equation}
According to (\ref{JJ}), the supercharges (\ref{Q12}) of
the superextended PT system (\ref{Hsusy}) correspond to the
trivial, spin integrals $\sigma_1$ and $\sigma_2$ of the
free particle system (\ref{Hsusy0}). Other,  ``dressed"
integrals can be found in a similar way. They  are
displayed in Table \ref{T1}, where  the integrals $J$ for
$\mathcal{H}_0$ and corresponding dressed integrals
$\tilde{J}$ for $\mathcal{H}$ are shown, respectively, in
the first and the second rows.
\begin{table}[ht]
\caption{Undressed (free particle), $J$, and dressed (PT),
$\tilde{J}$, integrals} \label{T1}
\begin{center}\renewcommand{\arraystretch}{1.6}
\begin{tabular}{|c|c|c|c|c|c|c|c|c|c|c|c|c|}
  \hline
  % after \\: \hline or \cline{col1-col2} \cline{col3-col4} ...
  %Integral for $\mathcal{H}_0$ &
  $%\mathds
  {1}$ & $\mathcal{H}_0$ &  $\sigma_3$ & $\sigma_1$ & $\sigma_2$ & $p$ &
  $s_1$ &
  $s_2$ & $\mathcal{R}\sigma_1$ & $\mathcal{T}\sigma_1$ &
  $\mathcal{R}\mathcal{T}$
  & $\mathcal{R}$
  &  $-i\mathcal{R}\sigma_2s_1$
  \\
  \hline
  %Dressed integral for $\mathcal{H}$ &
  $\mathcal{H}$ & $\mathcal{H}^2$ & $\sigma_3\mathcal{H}$ & $Q_1$ & $Q_2$ &
  $\mathcal{P}_1$
  & $S_1\mathcal{H}$ & $S_2\mathcal{H}$ &
  $-\mathcal{R}\sigma_1\mathcal{H}$ &
  $\mathcal{T}\sigma_1\mathcal{H}$ &
  $-\mathcal{R}\mathcal{T}\mathcal{H}$
  & $\mathcal{Q}$
  &  $\mathcal{S}\mathcal{H}$
  \\
  \hline
  %\label{Tab1}
\end{tabular}
\end{center}
\end{table}

We have introduced  the following notations,
\begin{equation}\label{int1}
    p=-i\frac{d}{dx}\,,\quad
    s_1=p\sigma_2-\coth{2\tau}\,\cdot\sigma_1\,,\quad
    s_2=i\sigma_3 s_1\,,
\end{equation}
\begin{equation}\label{Lax}
    \mathcal{P}_1=-i\left(%
\begin{array}{cc}
  Z_\tau & 0 \\
  0 & Z_{-\tau} \\
\end{array}%
\right),\qquad
Z_{\tau}=A_\tau\frac{d}{dx}A^\dagger_{\tau}\,,
\end{equation}
\begin{equation}\label{SS}
    S_1=\left(%
\begin{array}{cc}
  0 & {X}^\dagger_\tau \\
  X_\tau & 0 \\
\end{array}%
\right),\qquad S_2=i\sigma_3S_1\,,
\end{equation}
\begin{equation}\label{X}
    {X}_\tau=\frac{d}{dx}-\Delta_\tau(x)\,,\qquad
    {X}^\dagger_\tau=-{X}_{-\tau}\,,
\end{equation}
\begin{equation}\label{G12}
    \mathcal{Q}=\left(%
\begin{array}{cc}
  \mathcal{R}{Y}_\tau & 0 \\
  0 & \mathcal{R}{Y}_{-\tau} \\
\end{array}%
\right),\qquad
%\end{equation}
%\begin{equation}\label{G1}
    \mathcal{S}=\left(%
\begin{array}{cc}
  \mathcal{R}{X}_{\tau} & 0 \\
  0 & \mathcal{R}{X}_{-\tau} \\
\end{array}%
\right),
\end{equation}
where
\begin{equation}\label{Del}
    \Delta_\tau(x)=\tanh (x-\tau)-\tanh(x+\tau)+\coth
    2\tau\,.
\end{equation}
Function $\Delta_\tau(x)$, that appears in the structure of
the first order operator $X_\tau$,  has the properties
\begin{equation}\label{Del+-}
    \Delta_{\tau}(-x)=\Delta_\tau(x)\,,\qquad
    \Delta_{-\tau}(x)=-\Delta_\tau(x)\,,
\end{equation}
and satisfies the Riccati equation of the form
\begin{equation}\label{DelRic}
    \Delta_\tau^2(x)+\Delta'_\tau(x)=2(\tanh^2(x+\tau)-1)+\coth^2{2\tau}\,.
\end{equation}
Eq. (\ref{DelRic}) is based  on the identity
\begin{equation}\label{tanhadd}
   1-\tanh(x+\tau)\tanh(x-\tau)+\coth{2\tau}
   \big(\tanh(x-\tau)-\tanh(x+\tau)\big)=0\,,
\end{equation}
which is the addition formula for the function $\tanh u$.

To find a map $s_a\rightarrow S_a\mathcal{H}$, $a=1,2$, the
identities
\begin{equation}\label{AX}
    A_{\tau} \left(\frac{d}{dx}+\coth{2\tau}\right)=
    X_{-\tau} A_{-\tau}\,,\qquad
    \left(\frac{d}{dx}-\coth{2\tau}\right)A^\dagger_{\tau}=
    A^\dagger_{-\tau}X_{\tau}\,
\end{equation}
have been employed. Using these identities and those
obtained from them by the change $\tau\rightarrow -\tau$,
we find that the first order differential operators
$X_\tau$ and ${X}_{\tau}^\dagger$, from which the integrals
$S_1$ and $S_2$ are composed, are also the intertwining
operators,
\begin{equation}\label{Xinter}
    {X}_{\tau} H_{\tau}=H_{-\tau}{X}_{\tau}\,,\qquad
     H_{\tau}{X}_{\tau}^\dagger=X^\dagger_{\tau}H_{-\tau}\,,
\end{equation}
cf. (\ref{HH}), (\ref{Y}).

There are also the intertwining relations
\begin{equation}\label{RIX}
    \mathcal{R}X_\tau=-X_{-\tau}\mathcal{R}\,,\qquad
       \mathcal{R}Y_\tau=Y_{-\tau} \mathcal{R}\,,\qquad
       \mathcal{R}Z_\tau=-Z_{-\tau} \mathcal{R}\,,\qquad
\end{equation}
\begin{equation}\label{RIY}
     \mathcal{T}X_\tau=X_{-\tau}\mathcal{T}\,,\qquad
    \mathcal{T}Y_\tau=Y_{-\tau}\mathcal{T}\,,\qquad
    \mathcal{T}Z_\tau=Z_{-\tau}\mathcal{T}\,.
\end{equation}

The first, $X$, and the second, $Y$, order differential
operators intertwine, in turn, not only the Hamiltonians
$H_\tau$ and $H_{-\tau}$, but also the third order
operators $Z_\tau$ and $Z_{-\tau}$,
\begin{equation}\label{XinterZ}
    X_\tau Z_\tau=Z_{-\tau}X_\tau\,,\qquad Z_\tau
    X_{-\tau}=X_{-\tau}Z_{-\tau}\,,\qquad
%\end{equation}
%\begin{equation}\label{YinterZ}
    Y_\tau Z_\tau=Z_{-\tau}Y_\tau,\qquad
    Z_{\tau}Y_{-\tau}=Y_{-\tau}Z_{-\tau}\,.
\end{equation}

The operators $X$, $Y$ and $Z$ satisfy the identities
\begin{equation}\label{XYZprod1}
    -X_{-\tau}X_\tau =H_\tau+C_{2\tau}^2-1\,,\qquad
    Y_{-\tau}Y_\tau=H_\tau^2\,,\qquad
    -Z_\tau^2=H_\tau^2(H_\tau-1)\,,
\end{equation}
\begin{equation}\label{XYZprod2}
    X_{-\tau}Y_\tau=Z_\tau+C_{2\tau} H_\tau\,,\qquad
    X_\tau Z_\tau= -C_{2\tau} X_\tau H_\tau -Y_\tau
    (H_\tau +C_{2\tau}^2-1)\,,
\end{equation}
\begin{equation}\label{XYZprod2+}
    Y_\tau Z_\tau=X_\tau H_\tau^2+C_{2\tau} Y_\tau H_\tau,
\end{equation}
where
\begin{equation}\label{Ctau}
    \mathcal{C}_{2\tau}=\coth 2\tau.
\end{equation}
Other relations we shall need are obtained from them by
Hermitian conjugation, with taking into account the
relations $X_\tau^\dagger=-X_{-\tau}$,
$Y_\tau^\dagger=Y_{-\tau}$, $Z_\tau^\dagger=-Z_\tau$, as
well as by the change $\tau\rightarrow -\tau$.

According to (\ref{JJ}) with $D=A_\tau$, $H=H_0$ and
$\tilde{H}=H_\tau$, the PT relations
(\ref{XYZprod1})--(\ref{XYZprod2+}) are just the dressed
free particle identities
\begin{equation}\label{XYZprod10}
    -\left(ip+\mathcal{C}_{2\tau}\right)
    \left(ip-\mathcal{C}_{2\tau}\right)=
    H_0+\mathcal{C}_{2\tau}^2-1\,,\quad
    A^\dagger_{-\tau}A_{-\tau}=H_0\,,\quad
    pH_0p=H_0(H_0-1)\,,
\end{equation}
\begin{equation}\label{XYZprod20}
    ip+\mathcal{C}_{2\tau}
    =\left(ip\right)+
    \left(\mathcal{C}_{2\tau}\right)\,,\quad
    -\left(ip-
    \mathcal{C}_{2\tau}\right)ip=
    \left(\mathcal{C}_{2\tau}\left(ip-\mathcal{C}_{2\tau}\right)
    +\left(H_0+\mathcal{C}_{2\tau}^2-1\right)\right)\,,
\end{equation}
\begin{equation}\label{XYZprod20+}
    H_0ip=\left(
    ip-\mathcal{C}_{2\tau}\right)H_0
    +\mathcal{C}_{2\tau}\,
    H_0\,.
\end{equation}

\vskip0.1cm

Return now to the information presented in Table \ref{T1}.
In addition to $\sigma_3$, $Q_1$ and $Q_2$, the  operators
$\mathcal{P}_1$, $S_1$, $S_2$, $\mathcal{Q}$, $\mathcal{S}$,
$\mathcal{R}\sigma_1$, $\mathcal{T}\sigma_1$, and
$\mathcal{R}\mathcal{T}$ are identified as
Hermitian integrals of motion of the super-extended system
$\mathcal{H}$. The integrals $\sigma_3$, $Q_1$, $Q_2$,
$S_1$, $S_2$ and $\mathcal{P}_1$ are local, while the
$\mathcal{R}$ and the integrals which include it in their
structure are nonlocal in $x$ operators. Curiously, the
nontrivial nonlocal integral $\mathcal{Q}$ is just the
dressed parity integral $\mathcal{R}$ of the free particle
system.

It is known that any $n$-gap quantum mechanical periodic or
non-periodic system possesses a nontrivial Lax integral of
the odd order $(2n+1)$ \cite{FiniteGap}. In the present case of
the one-gap PT self-isospectral system $\mathcal{H}$, the
dressed momentum operator, $\mathcal{P}_1$, is (up to the
numerical factor $-i$) the third order Lax operator. The
intertwining relations (\ref{XinterZ})
mean that the Lax integral $\mathcal{P}_1$ commutes also with
the first, $S_a$, and the second, $Q_a$, order integrals of
the system $\mathcal{H}$.

The diagonal nature of the integrals (\ref{G12}) means that
in addition to the local integral $Z_{\tau}$ (Lax
operator), the one-gap PT \emph{subsystem} $H_{\tau}$ has
also nontrivial nonlocal integrals of motion,
$\mathcal{R}Y_\tau$ and $\mathcal{R}X_\tau$, see below.

For the self-isospectral system $\mathcal{H}$, we have the
integrals of motion $\sigma_3$, $\mathcal{R}\sigma_1$,
$\mathcal{T}\sigma_1$, and those obtained from them by a
composition, $\mathcal{R}\sigma_2$, $\mathcal{T}\sigma_2$,
 $\mathcal{R}\mathcal{T}\sigma_3$, $\mathcal{R}\mathcal{T}$.
The square of each of these seven Hermitian integrals
equals one, and any of them may be chosen as a grading
operator $\Gamma$. All the integrals from this set which
include in their structure reflection operators
$\mathcal{R}$ and $\mathcal{T}$, except the integral
$\mathcal{RT}$, may be obtained from the integral
$\sigma_3$ by a unitary transformation,
\begin{equation}\label{UnRI12}
    U_a(r)\sigma_3 U_a^\dagger(r)=r\sigma_a , \quad
    {\rm where} \quad
    U_a(r)=\frac{1}{\sqrt{2}}(\sigma_3 +\sigma_a r),\quad
    a=1,2,\quad r=\mathcal{R},\mathcal{T},
\end{equation}
\begin{equation}\label{UnRI3}
    U(\mathcal{R},\mathcal{T})\sigma_3
    U^\dagger(\mathcal{R},\mathcal{T})=
    \mathcal{RT}\sigma_3,\qquad
     U(\mathcal{R},\mathcal{T})=U_1(\mathcal{R})U_1(\mathcal{T}),
\end{equation}
\begin{equation}\label{U=U}
    U_a(r)=U_a^\dagger(r),\qquad
    U(\mathcal{R},\mathcal{T})=U^\dagger(\mathcal{R},\mathcal{T}),
    \qquad
    U_a^2(r)=U^2(\mathcal{R},\mathcal{T})=1.
\end{equation}
There exists no unitary transformation that would relate
$\mathcal{RT}$ with  $\sigma_3$, or with any other integral
from this set. Since any of the four unitary operators
$U_a(r)$, being composed from the integrals of motion,
commutes with $\mathcal{H}$, the latter is invariant under
any of five unitary transformations generated by $U_a(r)$
and $U(\mathcal{R},\mathcal{T})$.

With the listed above algebraic identities and intertwining
relations, we find that the basic \emph{local} integrals of
the first, second and third orders, $S_1$, $Q_1$ and
$\mathcal{P}_1$, are related between themselves and
Hamiltonian $\mathcal{H}$,
\begin{equation}\label{squares}
    S_1^2=\mathcal{H}+\mathcal{C}_{2\tau}^2-1\,,\qquad
    Q_1^2=\mathcal{H}^2\,,\qquad
    \mathcal{P}_1^2=\mathcal{H}^2(\mathcal{H}-1)\,,
\end{equation}
\begin{equation}\label{QS}
    S_1Q_1=-i\sigma_3\mathcal{P}_1-
    \mathcal{C}_{2\tau}\mathcal{H}\,,\qquad
    Q_1S_1=i\sigma_3\mathcal{P}_1
    -\mathcal{C}_{2\tau}\mathcal{H}\,,
\end{equation}
\begin{equation}\label{PS}
    \mathcal{P}_1S_1=S_1\mathcal{P}_1=-i\sigma_3\left(
    Q_1\left(\mathcal{H}+\mathcal{C}_{2\tau}^2-1\right)
    +\mathcal{C}_{2\tau}\mathcal{H}S_1\right)\,,
\end{equation}
\begin{equation}\label{PQ}
    \mathcal{P}_1Q_1=Q_1\mathcal{P}_1=i\sigma_3
    \left(S_1\mathcal{H}^2
    +\mathcal{C}_{2\tau}\mathcal{H}Q_1\right)\,.
\end{equation}

 These identities reproduce modulo $\mathcal{H}$ the
polynomial relations between the corresponding integrals
$s_1$, $\sigma_1$ and $p$ of the free particle system
$\mathcal{H}_0$.

\section{Eigenstates of $\mathcal{P}_1$, $Q_1$ and $S_1$}

{}In accordance with (\ref{PS}) and (\ref{PQ}), there
exists a common  basis for the integral $\mathcal{P}_1$ and
for one of the integrals $Q_1$ or $S_1$. The two sets of
corresponding eigenstates can be presented in a unified
form,
\begin{equation}\label{Leig01}
    \Psi^{0,1}_{\Lambda,+}=
    \left(%
\begin{array}{c}
  \Psi^{0,1}(x+\tau) \\
  \epsilon^{0,1}_\Lambda\Psi^{0,1}(x-\tau) \\
\end{array}%
\right),\qquad
    \Psi^{0,1}_{\Lambda,-}=\sigma_3\Psi^{0,1}_{\Lambda,+}\,,
\end{equation}
\begin{equation}\label{Leig01HP}
    \mathcal{H}\Psi^{0}_{\Lambda,\epsilon}=0,\qquad
    \mathcal{H}\Psi^{1}_{\Lambda,\epsilon}=\Psi^{1}_{\Lambda,\epsilon}\,,
    \qquad
    \mathcal{P}_1\Psi^{0,1}_{\Lambda,\epsilon}=0\,,\qquad
    \epsilon=\pm\,,
\end{equation}
\begin{equation}\label{Leigk+-}
    \Psi^{\pm k}_{\Lambda,+}=
    \left(%
\begin{array}{c}
  \psi^{\pm k}(x+\tau) \\
  e^{\pm i\varphi_{{}_\Lambda}(k,\tau)} \psi^{\pm k}(x-\tau) \\
\end{array}%
\right),\qquad
    \Psi^{\pm k}_{\Lambda,-}=\sigma_3\Psi^{\pm
    k}_{\Lambda,+}\,,
\end{equation}
\begin{equation}\label{Leigk+-HP}
    \mathcal{H}\Psi^{\pm k}_{\Lambda,\epsilon}=(1+k^2)\,
    \Psi^{\pm k}_{\Lambda,\epsilon}\,,\qquad
   \mathcal{P}_1\Psi^{\pm k}_{\Lambda,\epsilon}=
   \pm k(1+k^2)\,
    \Psi^{\pm k}_{\Lambda,\epsilon}\,,
\end{equation}
where $\Lambda=Q_1$ or  $S_1$, and $\Psi^{0}$, $\Psi^1$ and
$\psi^{\pm k}$ are the functions defined in (\ref{specsin})
and (\ref{specdoub}),
\begin{equation}\label{phiL}
    \epsilon^{0,1}_{Q_1}=-\epsilon^{0,1}_{S_1}=+1\,,\qquad
    e^{i\varphi_{{}_{Q_1}}(k,\tau)}=e^{2ik\tau}\,,\qquad
    e^{
    i\varphi_{{}_{S_1}}(k,\tau)}=e^{2ik\tau+i\theta(k,\tau)}\,,
\end{equation}
\begin{equation}\label{theta}
    e^{i\theta(k,\tau)}=e^{-i\theta(-k,\tau)}=
    -e^{-i\theta(k,-\tau)}=
    \frac{ik-\mathcal{C}_{2\tau}}{\sqrt{k^2+\mathcal{C}_{2\tau}^2}}\,,
\end{equation}
\begin{equation}\label{Qeig01}
    Q_1\Psi^{0}_{Q_1,\epsilon}=0,\qquad
    Q_1\Psi^{1}_{Q_1,\epsilon}=\epsilon \Psi^{1}_{Q_1,\epsilon}\,,\qquad
    Q_1\Psi^{\pm k}_{Q_1,\epsilon}=\epsilon(1+k^2)\Psi^{\pm
    k}_{Q_1,\epsilon}\,,
\end{equation}
\begin{equation}\label{Seig01}
    S_1\Psi^{0}_{S_1,\epsilon}=\epsilon\,
    \frac{1}{\sinh{2\tau}}\Psi^{0}_{S_1,\epsilon},\quad
    S_1\Psi^{1}_{S_1,\epsilon}=\epsilon\, \mathcal{C}_{2\tau}
    \Psi^{1}_{S_1,\epsilon}\,,\quad
    S_1\Psi^{\pm k}_{S_1,\epsilon}=\epsilon
    \sqrt{k^2+\mathcal{C}_{2\tau}^2}\,\Psi^{\pm
    k}_{S_1,\epsilon}\,.
\end{equation}

Antidiagonal operators $Q_1$ and $S_1$ anticommute with
$\sigma_3$, and multiplication by $\sigma_3$ changes their
eigenstates into eigenstates with an eigenvalue of the
opposite sign. {}From these relations we see that any pair
of mutually commuting operators, $(\mathcal{P}_1,Q_1)$ or
$(\mathcal{P}_1,S_1)$, provides the complete information
about the Hamiltonian eigenstates.

Hamiltonian $\mathcal{H}$ does not distinguish the
eigenstates different in index $\epsilon$, and does not
separate the states with index $+k$ and $-k$ in the
continuous part of the spectrum. Lax integral $\mathcal{P}_1$
distinguishes the states with index $+k$ and $-k$, but is
insensitive to the index $\epsilon$, and does not separate
the doublet states of energy $E=0$ and $E=1$, annihilating
all the corresponding four states
$\Psi^{0,1}_{\Lambda,\pm}$. The integrals $Q_1$ and
$S_1$ distinguish the states with $E=0$ and $E=1$, detect a
difference between the states with $\epsilon=+$ and
$\epsilon=-$, but do not separate the states with index
$+k$ and $-k$. The only integral that detects by its
eigenvalues a displacement $2\tau$ between the two
subsystems is $S_1$. Its eigenvalues of the bound states
$\Psi^0_{S_1,\epsilon}$, as well as of the continuous
spectrum eigenstates blow up, however, in the limit of a
zero shift, $\tau\rightarrow 0$.

The spectrum of $Q_1$ coincides with that of  the operator
$\sigma_3\mathcal{H}$. This is not casual\,: these two
operators are Darboux-dressed form of the free particle
integrals $\sigma_1$ and $\sigma_3$, which can be related
by a unitary transformation.

As follows from Table \ref{T2} below, the \emph{nonlocal}
integral $\mathcal{R}\sigma_1$ commutes with both integrals
$Q_1$ and $S_1$. Acting on their eingestates, it detects
the nontrivial relative phases between the upper and lower
components of the eigenstates,
\begin{equation}\label{RsQeig1}
    \mathcal{R}\sigma_1\Psi^{0}_{Q_1,\epsilon}=
    \epsilon \Psi^{0}_{Q_1,\epsilon}\,,\qquad
    \mathcal{R}\sigma_1\Psi^{1}_{Q_1,\epsilon}=
    -\epsilon\Psi^{1}_{Q_1,\epsilon}\,,
\end{equation}
\begin{equation}\label{RsSeig1}
    \mathcal{R}\sigma_1\Psi^{0}_{S_1,\epsilon}=
    -\epsilon \Psi^{0}_{S_1,\epsilon}\,,\qquad
    \mathcal{R}\sigma_1\Psi^{1}_{S_1,\epsilon}=
    \epsilon\Psi^{1}_{S_1,\epsilon}\,,
\end{equation}
\begin{equation}\label{RsLpmkeig}
    \mathcal{R}\sigma_1\Psi^{\pm k}_{\Lambda,\epsilon}=
    -\epsilon e^{\pm i\varphi_{{}_{\Lambda}}(k,\tau)}
    \Psi^{\mp
    k}_{\Lambda,\epsilon}\,.
\end{equation}
This can be compared with the action of another nonlocal
integral, $\mathcal{RT}$, that also changes index $+k$ for
$-k$ of the scattering states, but does not detect the
corresponding relative phases,
\begin{equation}\label{RTeig}
    \mathcal{RT}\Psi^{0}_{\Lambda,\epsilon}=
    \Psi^{0}_{\Lambda,\epsilon}\,,\quad
    \mathcal{RT}\Psi^{1}_{\Lambda,\epsilon}=
    -\Psi^{1}_{\Lambda,\epsilon}\,,\quad
    \mathcal{RT}\Psi^{\pm k}_{Q_1,\epsilon}=
    -\Psi^{\mp k}_{Q_1,\epsilon}\,,\quad
    \mathcal{RT}\Psi^{\pm k}_{S_1,\epsilon}=
    -\Psi^{\mp k}_{S_1,-\epsilon}\,.
\end{equation}
The difference in the two last relations in (\ref{RTeig})
originates from
commutativity of $\mathcal{RT}$ with $Q_1$ and its
anticommutativity with $S_1$, see Table \ref{T2}.

Finally, we note that though $Q_1$ and $S_1$ do not
(anti)commute and each of them does  not distinguish
indexes $+k$ and $-k$ of the states in the continuous part
of the spectrum, according to (\ref{QS}), the Lax integral
and Hamiltonian are reconstructed from them,
\begin{equation}\label{PSQrestore}
    \frac{i}{2}\sigma_3[S_1,Q_1]=\mathcal{P}_1\,,\qquad
    -\frac{1}{2\mathcal{C}_{2\tau}}\{S_1,Q_1\}=\mathcal{H}\,.
\end{equation}
Similarly, each pair of the integrals $(\mathcal{P}_1,S_1)$
or $(\mathcal{P}_1,Q_1)$ allows us to reconstruct the third
operator, respectively, $Q_1$ or $S_1$, see Eqs.
(\ref{PS}), (\ref{PQ}).

This information constitutes a part of a nonlinear
superalgebraic structure of the system, which we discuss in
the next Section.

\section{Nonlinear supersymmetries of self-isospectral PT
system}

For the grading operator $\Gamma=\sigma_3$, the
anti-diagonal \emph{local} integrals $Q_a$ and $S_a$ are
identified as fermionic operators, while the diagonal
integrals $\mathcal{P}_1$ and
$\mathcal{P}_2=\sigma_3\mathcal{P}_1$ should be treated as
bosonic generators of the superalgebra. The nonlinear
superalgebraic relations (\ref{QQ}) are extended then, in
correspondence with Eqs. (\ref{squares})--(\ref{PQ}), to
the nonlinear superalgebra
\begin{equation}\label{SSsusy}
    \{S_a,S_b\}=2\delta_{ab}\left(\mathcal{H}+\mathcal{C}_{2\tau}^2-1
    \right)\,,\qquad
     \{Q_a,Q_b\}=2\delta_{ab}\mathcal{H}^2\,,
\end{equation}
\begin{equation}\label{SQsusy}
    \{S_a,Q_b\}=-2\delta_{ab}\mathcal{C}_{2\tau}\mathcal{H}-2
    \epsilon_{ab}\mathcal{P}_1\,,
\end{equation}
\begin{equation}\label{PSsusy}
    [\mathcal{P}_2,S_a]=-2i\left(\left(\mathcal{H}+\mathcal{C}_{2\tau}^2-1\right)Q_a+
    \mathcal{C}_{2\tau}\mathcal{H}S_a\right),
\end{equation}
\begin{equation}\label{PQsusy}
    [\mathcal{P}_2,Q_a]=2i\left(\mathcal{H}^2S_a+\mathcal{C}_{2\tau}
    \mathcal{H}Q_a\right),
\end{equation}
\begin{equation}\label{Pcentr}
    [\mathcal{P}_1,S_a]=[\mathcal{P}_1,Q_a]=[\mathcal{P}_1,\mathcal{P}_a]=0\,,
\end{equation}
\begin{equation}\label{sigmaQSP}
    [\sigma_3,Q_a]=-2i\epsilon_{ab}Q_b\,,\qquad
    [\sigma_3,S_a]=-2i\epsilon_{ab}S_b\,,\qquad
    [\sigma_3,\mathcal{P}_a]=0\,,
\end{equation}
in which the Lax operator $\mathcal{P}_1$ plays
the role of the central charge.

The last relation from (\ref{squares}) does not show in the
(anti)commutation relations. It displays, however, in the
superalgebraic relations of the local integrals for any
other choice of the grading operator since then at least
one of the two integrals $\mathcal{P}_a$ is identified as
an odd, fermionic operator. The $\Z_2$-parity $\zeta=\pm$,
$\Gamma \mathcal{A}\Gamma=\zeta\mathcal{A}$, of the local
and some nonlocal integrals for all the choices of the
grading operator $\Gamma$ is shown in Table 2, where Eqs.
(\ref{RIX}), (\ref{RIY}) and relations
\begin{equation}\label{sigma22}
    \sigma_1\left(%
\begin{array}{cc}
  a & b \\
  c & d \\
\end{array}%
\right)\sigma_1=\left(%
\begin{array}{cc}
  d & c \\
  b & a \\
\end{array}%
\right),\qquad
\sigma_2\left(%
\begin{array}{cc}
  a & b \\
  c & d \\
\end{array}%
\right)\sigma_2=\left(%
\begin{array}{cc}
  d & -c \\
  -b & a \\
\end{array}%
\right)
\end{equation}
for $2\times 2$ matrices have been used~\footnote{Though
Hermitian unitary operators (\ref{UnRI12}) and
(\ref{UnRI3}) also commute with the Hamiltonian
$\mathcal{H}$ and their square equals one, they do not
assign a definite $\Z_2$-parity to some  of the nontrivial
integrals listed in Table \ref{T2}.}.

\begin{table}[ht]
\caption{$\Z_2$-parity of the local and some nonlocal
integrals} \label{T2}
\begin{center}
\begin{tabular}{|c||c|c|c|c|c|c|c|c|c|c|c|c|}\hline
$\Gamma$ & $Q_1$ & $Q_2$ & $S_1$ & $S_2$ & $\mathcal{P}_1$
& $\mathcal{P}_2$ & $\sigma_3$ & $\mathcal{S}$ &
$\mathcal{Q}$ & $\sigma_3\mathcal{S}$ &
$\sigma_3\mathcal{Q}$ & $\mathcal{RT}$
\\\hline\hline
$\sigma_3$ & $-$ & $-$ & $-$
& $-$ & $+$ & $+$ & $+$
& $+$ & $+$ & $+$ & $+$ & $+$ \\[1pt]\hline
$\mathcal{R}\sigma_1$ & $+$ & $-$ & $+$
& $-$ & $-$ & $+$ & $-$
& $-$ & $+$ & $+$ & $-$ & $+$ \\[1pt]\hline
$\mathcal{T}\sigma_1 $ & $+$ & $-$ & $-$ & $+$ & $+$ & $-$
& $-$
& $+$ & $+$ & $-$ & $-$ & $+$ \\[1pt]\hline
$\mathcal{R}\sigma_2$ & $-$ & $+$ & $-$ & $+$ & $-$ & $+$ &
$-$
& $-$ & $+$ & $+$ & $-$ & $+$ \\[1pt]\hline
$\mathcal{T}\sigma_2$ & $-$ & $+$ & $+$ & $-$ & $+$ & $-$ &
$-$
& $+$ & $+$ & $-$ & $-$ & $+$ \\[1pt]\hline
$\mathcal{R}\mathcal{T}\sigma_3$ & $-$ & $-$ & $+$ & $+$ &
$-$ & $-$ & $+$
& $-$ & $+$ & $-$ & $+$ & $+$ \\[1pt]\hline
$\mathcal{R}\mathcal{T}$ & $+$ & $+$ & $-$ & $-$ & $-$ &
$-$ & $+$
& $-$ & $+$ & $-$ & $+$ & $+$ \\[1pt]\hline
\end{tabular}
\end{center}
\end{table}

Notice that for any choice of the grading operator, the
nonlocal integral $\mathcal{Q}$, like $\mathcal{H}$ and
$\mathcal{RT}$, is an even operator.

As another example, we display the nonlinear superalgebraic
relations satisfied by the \emph{local} integrals for the
choice $\Gamma=\mathcal{R}\mathcal{T}$,
\begin{equation}\label{RISP}
    \{S_a,S_b\}=2\delta_{ab}\left(\mathcal{H}+\mathcal{C}_{2\tau}^2-1
    \right)\,,
\end{equation}
\begin{equation}\label{RIPP}
     \{\mathcal{P}_1,\mathcal{P}_1\}=\{\mathcal{P}_2,\mathcal{P}_2\}=
     2\mathcal{H}^2(\mathcal{H}-1)\,,\qquad
     \{\mathcal{P}_1,\mathcal{P}_2\}=2\mathcal{H}^2(\mathcal{H}-1)\sigma_3\,,
\end{equation}
\begin{equation}\label{RISP+}
    \{S_a,\mathcal{P}_1\}=-2\epsilon_{ab}\left(
    (\mathcal{H}+\mathcal{C}_{2\tau}^2-1)Q_b+
    \mathcal{C}_{2\tau}\mathcal{H}S_b\right)\,,\qquad
     \{S_a,\mathcal{P}_2\}=0\,,
\end{equation}
\begin{equation}\label{RIQP}
    [Q_1,Q_2]=-2i\mathcal{H}^2\sigma_3\,,\qquad
    [Q_a,\mathcal{P}_1]=0\,,
\end{equation}
\begin{equation}\label{RIQS}
    [Q_a,S_b]=2i\left(\delta_{ab}\mathcal{P}_2+\epsilon_{ab}
    \mathcal{C}_{2\tau}\mathcal{H}\sigma_3\right)\,,
    \qquad
    [Q_a,\mathcal{P}_2]=-2i\left(\mathcal{H}^2S_a
    +\mathcal{C}_{2\tau}\,\mathcal{H} Q_a\right)\,,
\end{equation}
which have to be completed by Eq. (\ref{sigmaQSP}).

The even generator $\sigma_3$ appears  only in
(\ref{sigmaQSP}) in superalgebra with $\Gamma=\sigma_3$,
while for $\Gamma=\mathcal{RT}$ it is present also in the
(anti)commutation relations (\ref{RIPP}) and (\ref{RIQS}).
Another, essential difference between both superalgebras is
that in the second case the constant
$\mathcal{C}_{2\tau}=\coth{2\tau}$ anticommutes with the
grading operator $\mathcal{RT}$ and  has to be treated
there as  an \emph{odd } generator of the superalgebra.
With such interpretation, the anticommutator (\ref{RISP+})
and the commutators in (\ref{RIQS}) produce, respectively,
even and odd combinations of the generators. In the case
$\Gamma=\sigma_3$, the $\mathcal{C}_{2\tau}$ should be
treated as the even central charge. In both superalgebras,
the Hamiltonian $\mathcal{H}$ appears as a multiplicative
central charge, that makes them nonlinear. A picture is
similar for other choices of the grading operator shown in
Table \ref{T2}.

The supersymmetric structure of the self-isospectral
one-gap PT system generated by local integrals of motion
admits therefore different choices for the grading
operator; each corresponding form of the superalgebra is
centrally extended and nonlinear. According to Eq.
(\ref{squares}), (\ref{specsin}), only the integrals
$\mathcal{P}_a$ annihilate the singlet states of the
isospectral subsystems $H_\tau$ and $H_{-\tau}$. On the
other hand,  the integrals $S_a$ have an empty kernel,
while the $Q_a$, $a=1,2$, annihilate only the states of
zero energy. Having in mind that for any choice of the
grading operator at least two integrals from the set of the
four integrals $Q_a$ and $S_a$ are identified as fermionic
generators, we always have \emph{partially broken}
nonlinear supersymmetry, cf. this picture with that  of
supersymmetry in the systems with topologically nontrivial
Bogomolny-Prasad-Sommerfield states \cite{BPS}.

\vskip0.1cm

One can find a modification of the integrals $S_1$ and
$S_2$, which annihilate the doublet of the ground states of
the self-isospectral system, by combining them with the
(non-local in the shift parameter $\tau$)  integral
$\mathcal{T}\sigma_1$. We get it using the explicit form
(\ref{Leig01}) of
the zero energy eigenstates  of $S_1$,
\begin{equation}\label{Sover}
    \overline{S}_1=S_1+\frac{1}{\sinh{2\tau}}\mathcal{T}\sigma_1,\qquad
    \overline{S}_2=i\sigma_3\overline{S}_1,\qquad
    \overline{S}_a\Psi^{0\pm}_{S_1,\epsilon}=0\,.
\end{equation}

 The modified integrals $\overline{S}_a$,
$a=1,2$, are odd supercharges with respect to the both
choices of the grading operator, $\Gamma=\sigma_3$ and
$\Gamma=\mathcal{RT}$, which correspond to the both
discussed superalgebras. The price to pay, however,  is
that the integrals $\overline{S}_a$ are not only
\emph{non-local} in the shift parameter, but also are
\emph{non-Hermitian},
$\overline{S}_1^\dagger=S_1-\sinh^{-1}{2\tau}\mathcal{T}\sigma_1\neq
\overline{S}_1$, and similarly, $\overline{S}_2^\dagger\neq
\overline{S}_2$. We have used here the relation
$(\sinh^{-1}{2\tau}\,\mathcal{T})^\dagger=\mathcal{T}\sinh^{-1}{2\tau}=
-\sinh^{-1}{2\tau}\,\mathcal{T}$. The modified supercharges
$\overline{S}_1$ and $\overline{S}_1^\dagger$ satisfy,
particularly, the anticommutation relations~\footnote{Some
similar integrals for the not self-isospectral
supersymmetric PT systems were discussed in \cite{BPS} in
the context of shape invariance, see also \cite{CDP}.
Unlike the present case, however, the integrals considered in
\cite{BPS} do not anticommute with the corresponding
grading operator $\sigma_3$, and their treatment as
fermionic generators  in the superalgebraic relations is
not justified there.}
\begin{equation}\label{SSmodif}
    \{\overline{S}_1,\overline{S}_1\}=
    \{\overline{S}_1^\dagger,\overline{S}_1^\dagger\}=2\mathcal{H}\,,\qquad
    \{\overline{S}_1,\overline{S}_1^\dagger\}=
    2\left(\mathcal{H}+2(\mathcal{C}_{2\tau}^2-1)\right)\,.
\end{equation}

\section{Hidden supersymmetry of unextended one-gap PT system}

We show here that the unextended, one-gap reflectionless PT
system is characterized by an exotic hidden supersymmetry,
which can be obtained from the supersymmetry of the
self-isospectral system $\mathcal{H}$ by  a nonlocal
Foldy-Wouthuysen transformation with a subsequent
reduction.

Indeed, the \emph{nonlocal} integrals (\ref{G12}) can be
got from the corresponding \emph{local} integrals by
applying a \emph{nonlocal} unitary transformation
\begin{equation}\label{U}
    \tilde{\mathcal{O}}=U_1(\mathcal{R})\mathcal{O}U^\dagger_1(\mathcal{R}),
    \qquad
    U_1(\mathcal{R})=\frac{1}{\sqrt{2}}(\sigma_3+\sigma_1\mathcal{R}),\qquad
    U^\dagger_1(\mathcal{R})=U_1(\mathcal{R}),\qquad
    U^2_1(\mathcal{R})=1\,,
\end{equation}
see Eq. (\ref{UnRI12}).
 This transformation does not change the form of the
self-isospectral Hamiltonian,
$\tilde{\mathcal{H}}=\mathcal{H}$, while
\begin{equation}\label{susytrans}
    \tilde{S_1}=\mathcal{S},\quad
    \tilde{S_2}=-S_2,\quad
    \tilde{Q}_1=\sigma_3\mathcal{Q},\quad
    \tilde{Q}_2=-Q_2,\quad
    \tilde{\mathcal{P}_1}=i\mathcal{R}\sigma_2\mathcal{P}_1,\quad
    \tilde{\mathcal{P}_2}=-\mathcal{P}_2,\quad
    \widetilde{\mathcal{T}\sigma_1}=\sigma_3\mathcal{RT},
\end{equation}
where $\mathcal{S}$ and $\mathcal{Q}$ are the nonlocal
integrals (\ref{G12}). The transformation (\ref{susytrans})
diagonolizes the supercharges $S_1$ and $Q_1$, and may be
treated as a kind of Foldy-Wouthuysen transformation. At
the same time, it does not change the diagonal form of the
integral $\mathcal{P}_2$.

The PT subsystem $H_\tau$, which is just a reduction of the
system $\mathcal{H}$ to the subspace $\sigma_3=+1$,  has
therefore one local and two nonlocal nontrivial integrals
\begin{equation}\label{pirho}
    \hat{\mathcal{P}}_1=-iZ_\tau\,,\qquad
    \hat{\mathcal{S}}_1=\mathcal{R}X_\tau\,,\qquad
    \hat{\mathcal{Q}}_1=\mathcal{R}Y_\tau\,.
\end{equation}
Making use of the intertwining relations
(\ref{RIX}), (\ref{RIY}) and relations (\ref{XYZprod1}),
(\ref{XYZprod2}), the Lax integral $\hat{\mathcal{P}}_1$ is
reconstructed from the integrals $\hat{\mathcal{S}}_1$ and
$\hat{\mathcal{Q}}_1$,
\begin{equation}\label{hatPQS}
    \hat{\mathcal{P}}_1=\frac{i}{2}\{\hat{\mathcal{S}}_1,
    \hat{\mathcal{Q}}_1\}\,,
\end{equation}
cf. (\ref{PSQrestore}). Similarly, $\hat{\mathcal{S}}_1$
($\hat{\mathcal{Q}}_1$) can be reconstructed from the
commutator of $\hat{\mathcal{Q}}_1$ ($\hat{\mathcal{S}}_1$)
with $\hat{\mathcal{P}}_1$.

Taking the integral
\begin{equation}\label{hatGam}
    \hat{\Gamma}=\mathcal{RT}\,,
\end{equation}
 $\hat{\Gamma}^2=1$,
$[H_\tau,\hat{\Gamma}]=0$,  as the grading operator, we
identify the integrals $\hat{\mathcal{P}}_1$ and
$\hat{\mathcal{S}}_1$ as odd, fermionic operators, while
$\hat{\mathcal{Q}}_1$ is identified as even, bosonic
operator. They can be supplied with two fermionic,
$\hat{\mathcal{P}}_2=i\hat{\Gamma}\hat{\mathcal{P}}_1$,
$\hat{\mathcal{S}}_2=i\hat{\Gamma}\hat{\mathcal{S}}_1$, and
one bosonic,
$\hat{\mathcal{Q}}_2=\hat{\Gamma}\hat{\mathcal{Q}}_1$,
integrals,
\begin{equation}\label{hat2}
    \hat{\mathcal{P}}_2=\mathcal{RT}Z_\tau\,,\qquad
    \hat{\mathcal{S}}_2=
    i\mathcal{T}X_\tau\,,\qquad
    \hat{\mathcal{Q}}_2=
    i\mathcal{T}Y_\tau\,.
\end{equation}
The nonlinear superalgebra generated by
$\hat{\mathcal{S}}_a$, $\hat{\mathcal{Q}}_a$,
$\hat{\mathcal{P}}_a$ and $H_\tau$ is
\begin{equation}\label{RISPhat1}
    \{\hat{\mathcal{S}}_a,\hat{\mathcal{S}}_b\}=
    2\delta_{ab}\left({H}_\tau+
    \mathcal{C}_{2\tau}^2-1\right)\,,\qquad
    \{\hat{\mathcal{P}}_a,\hat{\mathcal{P}}_b\}=
    2\delta_{ab}H_\tau^2(H_\tau-1)\,,
\end{equation}
\begin{equation}\label{RISPhat2}
     \{\hat{\mathcal{S}}_1,\hat{\mathcal{P}}_1\}=
     \{\hat{\mathcal{S}}_2,\hat{\mathcal{P}}_2\}=0\,,\qquad
    \{\hat{\mathcal{S}}_1,\hat{\mathcal{P}}_2\}=
     \{\hat{\mathcal{S}}_2,\hat{\mathcal{P}}_1\}=-2i\mathcal{C}_{2\tau}\hat{\mathcal{S}}_2H_\tau\,,
\end{equation}
\begin{equation}\label{RIQShat}
    [\hat{\mathcal{Q}}_a,\hat{\mathcal{Q}}_b]=0\,,\qquad
    [\hat{\mathcal{Q}}_1,\hat{\mathcal{S}}_a]=2i\hat{\mathcal{P}}_a\,,\qquad
    [\hat{\mathcal{Q}}_2,\hat{\mathcal{S}}_1]=
    -2\hat{\Gamma}\mathcal{C}_{2\tau} H_\tau\,,\qquad
    [\hat{\mathcal{Q}}_2,\hat{\mathcal{S}}_2]=
    -2i\mathcal{C}_{2\tau} H_\tau\,,
\end{equation}
\begin{equation}\label{RIQPhat}
    [\hat{\mathcal{Q}}_1,\hat{\mathcal{P}}_1]=-2i\left(
    H_\tau^2\hat{\mathcal{S}}_1+\mathcal{C}_{2\tau} H_\tau
    \hat{\mathcal{Q}}_1\right)
    \,,\quad
    [\hat{\mathcal{Q}}_1,\hat{\mathcal{P}}_2]=2\hat{\Gamma}\left(
    H_\tau^2\hat{\mathcal{S}}_1+\mathcal{C}_{2\tau} H_\tau
    \hat{\mathcal{Q}}_1\right)
    \,,\quad
    [\hat{\mathcal{Q}}_2,\hat{\mathcal{P}}_a]=0\,.
\end{equation}

As in the case of the superalgebra
(\ref{RISP})--(\ref{RIQS}), here the constant
$\mathcal{C}_{2\tau}$ anticommutes with the grading
operator $\hat{\Gamma}$, and  has to be treated as an
\emph{odd generator} of the superalgebra, that guarantees,
particularly, the correct Hermitian properties of the
(anti)commutation relations. For instance, for the r.h.s of
the last relation in (\ref{RISPhat2}) we have
$(-2i\mathcal{C}_{2\tau}\hat{\mathcal{S}}_2H_\tau)^\dagger=
+2iH_\tau\hat{\mathcal{S}}_2\mathcal{C}_{2\tau}=
-2i\mathcal{C}_{2\tau}\hat{\mathcal{S}}_2H_\tau$ in
correspondence with $(
\{\hat{\mathcal{S}}_1,\hat{\mathcal{P}}_2\})\dagger=
 \{\hat{\mathcal{S}}_1,\hat{\mathcal{P}}_2\}$, where
we have taken into account the Hermitian nature of the
involved integrals, and
$\hat{\mathcal{S}}_2\mathcal{C}_{2\tau}=-\mathcal{C}_{2\tau}
\hat{\mathcal{S}}_2$ due to the presence of the operator
$\mathcal{T}$ in the structure of the supercharge
$\hat{\mathcal{S}}_2$.

Notice also that the nature of superalgebra
(\ref{RISPhat1})--(\ref{RIQPhat}) of the \emph{hidden
supersymmetry} of the PT system $H_\tau$ has differences in
comparison with the both superalgebras of the self-isospectral
system discussed the previous Section. Particularly, the
operators $\hat{\mathcal{P}}_a$ ($\hat{\mathcal{Q}}_a$) are
odd (even) generators here in comparison with the even
(odd) nature of the integrals $\mathcal{P}_a$ ($Q_a$) in
the superalgebra (\ref{SSsusy})--(\ref{sigmaQSP}). Unlike
the superalgebras (\ref{SSsusy})--(\ref{sigmaQSP}) and
(\ref{RISP})--(\ref{RIQS}), the identity
$\hat{\mathcal{Q}}_a^2=H_\tau^2$ does not appear  in the
superalgebraic relations, cf. the second relation in
(\ref{SSsusy}) and the first relation in (\ref{RIQP}) with
taking into account  $Q_2=i\sigma_3 Q_1$.

\vskip0.1cm

Let us look how the basic supersymmetry generators
(\ref{pirho}) act on the states of the system. Before, we
note that though  the dependence on $\tau$ in the displaced
Hamiltonian $H_\tau$ and odd integral $\hat{\mathcal{P}}_1$
may be eliminated by the shift $x\rightarrow x-\tau$, the
parameter $-2\tau$ will still be present in the  structure
of the integrals $\hat{\mathcal{S}}_1$ and
$\hat{\mathcal{Q}}_1$, as well as in the grading operator
and in the invariant under such a shift superalgebraic
relations. Under such a shift, the reflection $\mathcal{R}$
with respect to $x=0$, that enters into the grading
operator $\hat{\Gamma}$, will be changed for the reflection
with respect to  $x=-\tau$.

On the other hand, though the non-shifted Hamiltonian
(\ref{PT}) is even while the Lax operator
$-iZ=-iA\frac{d}{dx}A^\dagger$ is odd with respect to the
reflection $\mathcal{R}$ in $x=0$ operator, the integrals
$\hat{\mathcal{S}}_1$ and $\hat{\mathcal{Q}}_1$ do not
possess a definite parity with respect to it~\footnote{The
integrals $-iZ$ and $\mathcal{R}Z$ generate together with
(\ref{H1}) the third order, nonlinear superalgebra with $\mathcal{R}$
identified as the grading operator, see \cite{CP1}.}.

The eigenstates of the shifted Hamiltonian $H_\tau$ we
denote here as in (\ref{specsin}), implying that their
argument is  $ x+\tau$. The eigenstates and eigenvalues of
the operators (\ref{pirho}) are
\begin{equation}\label{Zeig}
    \hat{\mathcal{P}}_1\Psi^0=\hat{\mathcal{P}}_1\Psi^1=0,\qquad
    \hat{\mathcal{P}}_1\psi^{\pm k}=\pm k (k^2+1)\psi^{\pm k}\,,
\end{equation}
\begin{equation}\label{Qeig}
    \hat{\mathcal{Q}}_1\Psi^0=0,\qquad
     \hat{\mathcal{Q}}_1
    \Psi^1= \Psi^1,\qquad
    \hat{\mathcal{Q}}_1\psi_{\hat{\mathcal{Q}}_1,\pm}^{k}=\pm (1+k^2)
    \psi_{\hat{\mathcal{Q}}_1,\pm}^{k}\,,
\end{equation}
\begin{equation}\label{Qeig+}
    \hat{\mathcal{S}}_1\Psi^0=-(\sinh{2\tau})^{-1}\Psi^0,\qquad
    \hat{\mathcal{S}}_1
    \Psi^1=\coth{2\tau}\Psi^1,\qquad
    \hat{\mathcal{S}}_1\psi_{\hat{\mathcal{S}}_1,\pm}^{k}=\pm
    \sqrt{k^2+\coth^2{2\tau}}\,
    \psi_{\hat{\mathcal{Q}}_1,\pm}^{k}\,,
\end{equation}
\begin{equation}\label{psiQS}
    \psi_{\hat{\mathcal{Q}}_1,\pm}^{k}=
    \frac{1}{2}\left(\psi^{+k}\pm e^{2ik\tau}\psi^{-k}\right),
    \qquad
    \psi_{\hat{\mathcal{S}}_1,\pm}^{k}=\frac{1}{2}\left(\psi^{+k}\mp e^{
    i\varphi_{{}_{S_{1}}}(k,\tau)}\psi^{-k}\right),
\end{equation}
cf. (\ref{Leig01})--(\ref{Seig01}), on which the grading
operator $\hat{\Gamma}$ acts as
\begin{equation}\label{Geig}
    \hat{\Gamma}\Psi^0=\Psi^0,\qquad
    \hat{\Gamma}\Psi^1=-\Psi^1,\qquad
    \hat{\Gamma}(\psi^{+k}\pm \psi^{-k})=\mp(\psi^{+k}\pm
    \psi^{-k}),
\end{equation}
\begin{equation}\label{GeigQS}
    \hat{\Gamma}\psi_{\hat{\mathcal{Q}}_1,\pm}^{k}=
    \mp e^{-2ik\tau}\psi_{\hat{\mathcal{Q}}_1,\pm}^{k},\qquad
    \hat{\Gamma}\psi_{\hat{\mathcal{S}}_1,\pm}^{k}=\mp e^{-i
    \varphi_{{}_{S_{1}}}(k,\tau)}\psi_{\hat{\mathcal{S}}_1,\mp}^{k}\,,
\end{equation}
cf. (\ref{RsQeig1})--(\ref{RsLpmkeig}), where
$\varphi_{{}_{S_{1}}}(k,\tau)$ is the phase defined by Eqs.
(\ref{phiL}), (\ref{theta}).

{}Like in the case of the extended, self-isospectral system
$\mathcal{H}$, the Hamiltonian $H_\tau$ distinguishes here
the singlet states $\Psi^0$ and $\Psi^1$, but does not
distinguish the doublet states $\psi^{\pm k}$. The Lax
operator $\hat{\mathcal{P}}_1$, instead, distinguishes the
doublet states, but does not distinguish the singlet
states\,: they both are its zero modes~\footnote{The third
state annihilated by the third order differential operator
$Z_\tau$ is $\cosh(x+\tau)$, which is a nonphysical state
of a free particle Hamiltonian (\ref{H0}) of zero
eigenvalue \cite{LPT}.}. The operator $\hat{\mathcal{Q}}_1$
distinguishes all the eigenstates of $H_\tau$, but it does
not detect a virtual here shift $\tau$, like the integral
$Q_1$ does not detect a real shift between the subsystems
of the extended self-isospectral system $\mathcal{H}$. Only
the integral $\hat{\mathcal{S}}_1$  distinguishes all the
states as well as detects a virtual here shift $2\tau$.

Unlike the integrals $\hat{\mathcal{P}}_1$ and
$\hat{\mathcal{Q}}_1$, the kernel of the integral
$\hat{\mathcal{S}}_1$ is empty\,: it annihilates a
nonphysical state~\footnote{The state (\ref{ftau}) is a
nonphysical eigenstate of $H_\tau$ of eigenvalue
$-\sinh^{-2}{2\tau}$. The second state annihilated by
$\hat{\mathcal{Q}}_1$ is $(\sinh 2x +
2x\cosh{2\tau})/\cosh(x+\tau)$, which is a nonphysical
eigenstate of $H_\tau$ of eigenvalue $0$.}
\begin{equation}\label{ftau}
    f_\tau(x)=\frac{\cosh(x-\tau)}{\cosh(x+\tau)}\,e^{x\coth 2\tau
    },
\end{equation}
in terms of which the operator (\ref{X}) is presented as
\begin{equation}\label{Xftau}
    X_\tau=f_\tau \frac{d}{dx}
    \frac{1}{f_\tau}=\frac{d}{dx}-(\ln f_\tau )'\,.
\end{equation}

The analogs of the modified supercharges (\ref{Sover}) here
are
\begin{equation}\label{redefS}
    \hat{\overline{\mathcal{S}}}_1= \hat{\mathcal{S}}_1
    +(\sinh{2\tau})^{-1}\hat{\Gamma}\,,\qquad
    \hat{\overline{\mathcal{S}}}_2=
    i\hat{\Gamma}\hat{\overline{\mathcal{S}}}_1\,,\qquad
    \{\hat{\Gamma},\hat{\overline{\mathcal{S}}}_a\}=0\,,\qquad
    \hat{\overline{\mathcal{S}}}_a\Psi^0=0\,,
\end{equation}
which are non-Hermitian, odd operators. Operators
$\hat{\overline{\mathcal{S}}}_1$  and
$\hat{\overline{\mathcal{S}}}_1^\dagger$ satisfy the
aticommutation relations
$\{\hat{\overline{\mathcal{S}}}_1,\hat{\overline{\mathcal{S}}}_1\}=
\{\hat{\overline{\mathcal{S}}}_1^\dagger,
\hat{\overline{\mathcal{S}}}_1^\dagger\}=2H_\tau$,
$\{\hat{\overline{\mathcal{S}}}_1,
\hat{\overline{\mathcal{S}}}_1^\dagger\}=2(H_\tau+2(\mathcal{C}_{2\tau}^2-1))$,
cf. (\ref{SSmodif}).

\section{Supersymmetry of  one-gap nonperiodic BdG system}

The self-isospectral supersymmetric structure  we have
discussed admits an interesting alternative interpretation
in terms of the associated one-gap, non-periodic
Bogoliubov-de Gennes system. In this Section we reveal a
set of (non)local integrals for the latter system, which
generate a non-linear supersymmetric structure to be of the
order eight in the BdG Hamiltonian.

Consider one of the local, first order integrals $S_a$ as a
$(1+1)D$ Dirac Hamiltonian. This corresponds to
the Bogoliubov-de Gennes system.
Depending on the
physical context, the function $\Delta_\tau$ plays a
role of an order parameter, a condensate, or a gap
function~\cite{Bish,SchTh,BasDun,CDP}.

For the sake of definiteness, we identify $S_1$ as a first
order Hamiltonian, $H_{BdG}=S_1$. It is a Darboux-dressed
form of the $(1+1)D$ Dirac Hamiltonian
$s_1=p\sigma_2-\coth{2\tau}$ of the free particle of mass
$m=\coth{2\tau}$. The energy gap $2m=2\coth{2\tau}$ in the
spectrum of the free Dirac particle transforms effectively by the
Darboux transformation (\ref{Dmat}) into the $x$-dependent
gap function 2$\Delta_\tau(x)$. The square of the free
Dirac particle Hamiltonian, (\ref{Hsusy0}), which is given
by the two copies of the free particle second order
Hamiltonian, transforms  into the Hamiltonian of the
self-isospectral PT system $\mathcal{H}$, whose eigenstates
are given by Eqs. (\ref{Leig01}), (\ref{Leigk+-}). Under
such a transformation, the mass parameter $m=\coth{2\tau}$
of the free particle system maps into a spatial shift
$2\tau$  of the two PT subsystems, $H_\tau$ and
$H_{-\tau}$.

Operator $\sigma_3$ anticommutes with the BdG Hamiltonian
$S_1$, and  plays a role of the energy reflection operator.
As follows from Table \ref{T2}, $H_{BdG}=S_1$ commutes with
$\mathcal{R}\sigma_1$, $\mathcal{T}\sigma_2$ and
$\mathcal{RT}\sigma_3$, any of which can be identified as
the grading operator for the BdG system.  These are
nonlocal in $x$ or $\tau$, or in both of them, trivial
integrals  of $H_{BdG}$. A nontrivial local BdG integral is
$\mathcal{P}_1$. The BdG Hamiltonian anticommutes with the
nonlocal integral $\mathcal{S}$ of the self-isospectral PT
system. The latter is just the Foldy-Wouthuysen
transformed, diagonal form of the BdG Hamiltonian $S_1$,
being a Darboux-dressed form of the operator
\begin{equation}\label{freeDFW}
    -i\mathcal{R}\sigma_2
    s_1=\mathcal{R}\left(-\frac{d}{dx}+
    \sigma_3\coth{2\tau}\right)\,,
\end{equation}
see Table \ref{T1}. Operator (\ref{freeDFW}) is the
Foldy-Wouthuysen transformed, diagonal form of the free
Dirac particle Hamiltonian $s_1$. The operator
\begin{equation}\label{S3sig}
    \sigma_3\mathcal{S}=(\mathcal{R}\sigma_1)S_1
\end{equation}
is then a nonlocal integral of $H_{BdG}$,
$[S_1,\sigma_3\mathcal{S}]=0$. $H_{BdG}$ still has one
more, nontrivial nonlocal integral. To identify it, we note
that with respect to $\mathcal{R}\sigma_1$,
$\mathcal{T}\sigma_2$ or $\mathcal{RT}\sigma_3$, the local
integral $\mathcal{P}_1$ is identified, respectively, as
the odd, even or odd operator, see Table \ref{T2}, while
the nonlocal integral $\sigma_3\mathcal{S}$ has,
respectively, even, odd and, once again, odd
$\Z_2$-parities. This means that in dependence on the
choice of the grading operator, we have to calculate either
commutator or anticommutator of these two integrals. We
find
\begin{equation}\label{PSBdG}
    \{\mathcal{P}_1,\sigma_3\mathcal{S}\}=0\,,\qquad
    [\mathcal{P}_1,\sigma_3\mathcal{S}]=-2i\mathcal{F}\,,
\end{equation}
where
\begin{equation}\label{FBdG}
    \mathcal{F}=-i\sigma_3\mathcal{S}\mathcal{P}_1=
    \mathcal{C}_{2\tau}\mathcal{S}
    (S_1^2-\mathcal{C}_{2\tau}^2+1)+\sigma_3\mathcal{Q}S_1^2
\end{equation}
is the third basic, nontrivial BdG nonlocal integral,
which is a Darboux dressed integral $\mathcal{R}\left(\frac{d}{dx}
-\mathcal{C}_{2\tau}\right)\frac{d}{dx}$ of the free Dirac particle.
The
$\Z_2$-parities of $\mathcal{F}$ with respect to
$\mathcal{R}\sigma_1$, $\mathcal{T\sigma}_2$ or
$\mathcal{RT}\sigma_3$ are, respectively, $-$, $-$, or $+$,
where the anticommutativity of $\mathcal{C}_{2\tau}$ with
$\mathcal{T}\sigma_2$ and $\mathcal{RT}\sigma_3$ has to be
taken into account.

Summarizing, for each of the three possible identifications
of the grading operator for the BdG system,
$\mathcal{R}\sigma_1$, $\mathcal{T}\sigma_2$, or
$\mathcal{RT}\sigma_3$,  one of the basic integrals,
respectively, $\sigma_3\mathcal{S}$, $\mathcal{P}_1$,  or
$\mathcal{F}$, is identified as the even generator, while
the two other integrals are identified each time as the
$\Z_2$-odd supercharges, see Table \ref{T3}.

\begin{table}[ht]
\caption{Possible grading operators and $\Z_2$-parities of
the basic BdG integrals} \label{T3}
\begin{center}
\begin{tabular}{|c||c|c|c|}\hline
$\Gamma$ & $\mathcal{P}_1$ & $\sigma_3\mathcal{S}$ &
$\mathcal{F}$
\\\hline\hline
$\mathcal{R}\sigma_1$ &  $-$ & $+$ & $-$  \\[1pt]\hline
$\mathcal{T}\sigma_2$ &  $+$ & $-$ & $-$ \\[1pt]\hline
$\mathcal{RT}\sigma_3$ & $-$ &  $-$ & $+$ \\[1pt]\hline
\end{tabular}
\end{center}
\end{table}

The set of the (anti)commutation
relations (\ref{PSBdG}) has to be extended then by
\begin{equation}\label{SF}
    \{\sigma_3\mathcal{S},\mathcal{F}\}=0\,,\qquad
    [\sigma_3\mathcal{S},\mathcal{F}]=2i\mathcal{P}_1S_1^2\,,
\end{equation}
\begin{equation}\label{PF}
    \{\mathcal{P}_1,\mathcal{F}\}=0\,,\qquad
    [\mathcal{P}_1,\mathcal{F}]=
    2i\left(S_1^2-\mathcal{C}_{2\tau}^2+1\right)^2
    \left(S_1^2-
    \mathcal{C}_{2\tau}^2\right)\sigma_3\mathcal{S}\,,
\end{equation}
\begin{equation}\label{P2S2}
    \mathcal{P}_1^2=\left(S_1^2-\mathcal{C}_{2\tau}^2+1\right)^2
    \left(S_1^2-\mathcal{C}_{2\tau}^2\right)\,,\qquad
    (\sigma_3\mathcal{S})^2=S_1^2\,,
\end{equation}
\begin{equation}\label{F2}
    \mathcal{F}^2=S_1^2\left(S_1^2-\mathcal{C}_{2\tau}^2\right)
    \left(S_1^2-\mathcal{C}_{2\tau}^2+1\right)^2\,.
\end{equation}

The action of the Lax integral $\mathcal{P}_1$ on the
eigenstates of the Hamiltonian $H_{BdG}=S_1$ is given by
Eqs. (\ref{Leig01HP}), (\ref{Leigk+-HP}), while the action
of the BdG integrals $\sigma_3\mathcal{S}$ and
$\mathcal{F}$ can be easily found by making use of Eqs.
(\ref{S3sig}), (\ref{FBdG}),
(\ref{Leig01})--(\ref{Seig01}), (\ref{RsSeig1}),
(\ref{RsLpmkeig}).

The spectrum of the $H_{BdG}=S_1$
is symmetric, $(-\infty,-\mathcal{E}_1)\cup -\mathcal{E}_0
\cup \mathcal{E}_0 \cup (\mathcal{E}_1,+\infty)$,
where $\mathcal{E}_0=\sinh^{-1}2\tau$,
$\mathcal{E}_1=\coth{2\tau}$. The eigenvalues
$\pm\mathcal{E}_0$ of the bound states, and
the eigenvalues $\pm\mathcal{E}_1$ of the edge states of the
continuous
parts of the spectrum are nondegenerate.
The continuous bands are separated
by the gap
$2\mathcal{E}_1=2\coth{2\tau}$, while
$\mathcal{E}_1^2-\mathcal{E}_0^2=1$.
All the corresponding
singlet states are annihilated by the integrals
$\mathcal{P}_1$ and $\mathcal{F}$,
while
$\sigma_3\mathcal{S}\Psi^0_{S_1,\epsilon}=
-\mathcal{E}_0\Psi^0_{S_1,\epsilon}$,
$\sigma_3\mathcal{S}\Psi^1_{S_1,\epsilon}=
\mathcal{E}_1\Psi^1_{S_1,\epsilon}\,$, cf.
Eq. (\ref{Seig01}).
The eigenstates and eigenvalues of $\sigma_3\mathcal{S}$
and $\mathcal{F}$ in the doubly degenerate
continuous parts of the spectrum are
given by
\begin{equation}\label{s3Seig}
    \sigma_3\mathcal{S}\left(
    \Psi^{+k}_{S_1,\epsilon}
    \pm e^{i\varphi_{{}_{S_1}}(k,\theta)}
    \Psi^{-k}_{S_1,\epsilon}\right)=
    \mp\sqrt{k^2+\mathcal{E}_1^2}
    \left(
    \Psi^{+k}_{S_1,\epsilon}
    \pm e^{i\varphi_{{}_{S_1}}(k,\theta)}
    \Psi^{-k}_{S_1,\epsilon}\right)\,,
\end{equation}
\begin{equation}\label{mathFeig}
    \mathcal{F}\left(
    \Psi^{+k}_{S_1,\epsilon}
    \pm ie^{i\varphi_{{}_{S_1}}(k,\theta)}
    \Psi^{-k}_{S_1,\epsilon}\right)=
    \pm k(1+k^2)\sqrt{k^2+\mathcal{E}_1^2}
    \left(
    \Psi^{+k}_{S_1,\epsilon}
    \pm ie^{i\varphi_{{}_{S_1}}(k,\theta)}
    \Psi^{-k}_{S_1,\epsilon}\right)\,.
\end{equation}
Since all the three basic integrals $\mathcal{P}_1$,
$\sigma_3\mathcal{S}$ and $\mathcal{F}$
commute with $\sigma_3$,
only the BdG Hamiltonian $S_1$ detects a difference
between the states with opposite values
of the low index $\epsilon$,
see Eq. (\ref{Seig01}).

Let us discuss the structure of the superalgebra of the BdG
system. The trivial integrals $\mathcal{R}\sigma_1$,
$\mathcal{T\sigma}_2$ and $\mathcal{RT}\sigma_3$ generate
between themselves the three-dimensional Clifford algebra,
i.e. the same algebra as the $\sigma_i$, $i=1,2,3$, do. For
any choice of the grading operator, two different basic odd
supercharges anticommute. The square of the each basic
integral,  $\sigma_3\mathcal{S}$, $\mathcal{P}_1$ and
$\mathcal{F}$,  is a polynomial in $H_{Bdg}^2=S_1^2$ of
the order, respectively, $1$, $3$ and $4$. A commutator of
any two basic integrals produces, modulo a certain
polynomial of  $H_{Bdg}^2=S_1^2$, a third integral. As a
result, for any choice of the grading operator, the
superalgebra has a somewhat similar structure to be a nonlinear
superalgebra, in which the $H_{Bdg}^2=S_1^2$ plays a
role of the multiplicative central charge.

As an explicit example, consider the case with
$\Gamma=\mathcal{R}\sigma_1$ chosen as the grading
operator, and denote
\begin{equation}\label{AABB}
    \mathcal{A}_1=\mathcal{P}_1\,,\qquad
    \mathcal{A}_2=i\Gamma\mathcal{A}_1\,,\qquad
    \mathcal{F}_1=\mathcal{F}_2\,,\qquad
    \mathcal{F}_2=i\Gamma\mathcal{F}_1\,,\qquad
    \mathcal{B}=\sigma_3\mathcal{S}\,,
\end{equation}
where $\mathcal{A}_a$ and $\mathcal{F}_a$, $a=1,2$, are
identified as the odd generators, while $\mathcal{B}$ is
the even generator. In these notations, a nonlinear
superalgebra of the one-gap, non-periodic BdG system can be
presented in a compact form,
\begin{equation}\label{superBdGalg1}
    \{\mathcal{A}_a,\mathcal{A}_b\}=2\delta_{ab}(S_1^2-
    \mathcal{C}_{2\tau}^2)
    (S_1^2-
    \mathcal{C}_{2\tau}^2+1)\,,\qquad
    \{\mathcal{A}_a,\mathcal{F}_b\}=0\,,
\end{equation}
\begin{equation}\label{superBdGalg2}
    \{\mathcal{F}_a,\mathcal{F}_b\}=2\delta_{ab}
    S_1^2\left(S_1^2-\mathcal{C}_{2\tau}^2\right)
    \left(S_1^2-\mathcal{C}_{2\tau}^2+1\right)^2\,,
\end{equation}
\begin{equation}\label{superBdGalg3}
    [\mathcal{B},\mathcal{A}_a]=2i\mathcal{F}_a\,,\qquad
    [\mathcal{B},\mathcal{F}_a]=2iS_1^2\mathcal{A}_a\,.
\end{equation}
This is a nonlinear superalgebra of the order eight in the BdG
Hamiltonian $H_{BdG}=S_1$.

\section{Discussion and outlook}

Our analysis of nonlinear supersymmetry
of the one-gap, reflectionless self-isospectral
P\"oschl-Teller system was based on a mirror symmetry
and a related Darboux dressing.

\vskip0.05cm
Mirror symmetry has
a twofold nature here. On the one hand, it is
generated by a spatial reflection, and by a reflection
of the parameter of a shift of the two
PT subsystems. On the other hand,
in the Darboux-Crum map between the two PT subsystems,
a free particle system appears as a virtual mirror,
by means of which the second order Darboux-Crum transformation
between the mutually shifted PT subsystems
factorizes into a sequence of the two first order Darboux
transformations.

\vskip0.05cm
In this construction, all the trivial and non-trivial
generators of the supersymmetry
of the self-isospectral PT system appear as the Darboux-dressed
integrals of the free spin-1/2  particle system described by
the second order Hamiltonian.
The first order, one-gap Bogoliubov-de Gennes system
associated with the self-isospectral second order PT system
is just a dressed free massive Dirac particle.
In such a picture, a mass parameter of the free Dirac particle
transforms effectively into a gap function of the BdG system.
The Dirac mass
maps into the parameter of the mutual shift (displacement)
of the two subsystems for the
second order self-isospectral PT system.

\vskip0.05cm
The key role in the exotic nonlinear supersymmetry
of the one-gap, reflectionless self-isospectral
P\"oschl-Teller system and the
associated first order Bogoliubov-de Gennes system is played
by the third order Lax operator, which is a diagonal
integral for the both systems, and is a dressed momentum
operator of the corresponding free particle systems.
In dependence on the choice of the grading operator,
for which there are, respectively, seven (PT) and three
(BdG) possibilities, it plays the role
of one of the even, or odd integrals of the motion.
Supersymmetric structures of the both, PT and BdG, systems,
include also two more \emph{basic} nontrivial
integrals, which provide a factorization of
the Lax operator, \emph{modulo}
a corresponding second, or first order Hamiltonian.

\vskip0.05cm The analysis, based on the mirror symmetry,
may be extended directly for the $n$-gap non-periodic case
by appropriate generalization of the Darboux-Crum
transformation. Our approach may also be applied to the
case of $n$-gap periodic, second order Lam\'e quantum
systems, and to the associated periodic BdG systems. Since
the one-gap P\"oschl-Teller potential may be achieved as a
limit case of the Lam\'e one with $n=1$, this,
particularly, will allow us to analyze in a new light a
connection between the algebraic structure associated with
the previously observed hidden supersymmetry in Lam\'e
systems \cite{CNP1,Tri} with the corresponding structure
studied here. All these generalizations will be presented
elsewhere.

\vskip0.15cm
%\newpage

 \noindent \textbf{Acknowledgements.}
The work of MSP has been partially supported by
 FONDECYT Grant 1095027, Chile and  by Spanish Ministerio de
 Educaci\'on under Project
SAB2009-0181 (sabbatical grant). LMN has been partially
supported by the Spanish Ministerio de Ciencia e
Innovaci\'on (Project MTM2009-10751) and Junta de Castilla
y Le\'on (Excellence Project GR224).

%%%%%%%%%%%%%%%%%%%%%%%%%%%%%%%%%%%%%%%%%%%%%%%%%%%%%%%%%%%%
%%%%%%%%%%%%%%%%%%%%%%%%%%%%%%%%%%%%%%%%%%%%%%%%%%%%%%%%%%%%

\end{document}